\def\selectedoptions{}
\ifx\empty\selectedoptions
  \def\selectedoptions{final}
\fi

\documentclass[
   \selectedoptions
  ]
  {aipproc}

\def\selectedlayoutstyle{6x9} 
\layoutstyle\selectedlayoutstyle

\SetInternalRegister\hbadness{8000} 

\newcommand\doingARLO[2][]{%
  \ifx\mmref\undefined #1\else #2\fi
}

\begin{document}

\title 
      []
      {Extracting $\mid V_{ub}\mid$ Using the Radiative Decay Data}


\author{Ira Z. Rothstein}{
  address={Dept. of Physics Carnegie Mellon University, Pittsburgh PA
  15213, USA},
  email={ira@cmuhep2.phys.cmu.edu},
  thanks={}
}


%
%

%

\copyrightyear  {2001}

\begin{abstract}
In this talk I review recent progress made in extracting $\mid V_{ub}
\mid$, within a systematic expansion, 
from the cut electron energy and hadronic mass spectra of inclusive
$B$ meson decays utilizing the data from radiative decays. It is shown
that an extraction is possible without modeling
the B meson structure function. I  discuss the issues involving
the assumptions of local duality in various extractions. I also
comment on the recent CLEO extraction of $V_{ub}$.
\end{abstract}

\date{\today}

\maketitle

\section{Introduction}
It is an unfortunate fact that experimental cuts can take a nice
clean theoretical prediction and turn it into a troublesome mess.
A perfect example of this scenario arises in the extraction of
$V_{ub}$ from inclusive $B$ decays. In principle this extraction
should be straightforward. One measures the inclusive
rate for semi-leptonic decays into non-charmed states and compares the
result
to the theoretical prediction for the total rate, which is under
good theoretical control \cite{CGG,SV1}. Of course, the snag is that
there is no simple way, at least at this time, to measure the total
inclusive rate to charmless states. Thus, some cut must be applied
to reject the charmed final states. Perhaps the simplest choice, from an
experimentalist viewpoint, is to cut on the electron energy,
rejecting all events with $E_l<(m_B^2-m_D^2)/(2m_B)$. This is the
oldest method for extracting $V_{ub}$. Unfortunately, the
theoretical prediction for the integrated cut spectrum is rather
complicated. 

The problem arises from the fact that the cut introduces a new scale
into the problem. Without the cut there are only two scales of
relevant physics, namely, the mass $m_B$  and
the QCD scale $\Lambda_{QCD}$. Suppose we cut on a scaled kinematic
variable, such as the electron energy $x_B=2E_l/m_B$. If we cut
near the endpoint, $x_B\simeq 1$, then the scale $(1-x)m_B$ can
introduce
a large  dimensionless ratio into the calculation $1/(1-x)$.
This can lead to power law, as well as 
logarithmic amplifications of what  normally would be small effects.   

The calculation of the total inclusive rate can be derived from
first principles \footnote{Even in the total rate for semi-leptonic
decays one still
has ``mild'' local-duality assumptions arising from the fact that
the contour approaches the real axis at a point. Thus this calculation
is technically not on the same theoretical footing as deep inelastic
scattering
} within a 
systematic expansion in $\alpha_s(m_b)$ and $\Lambda/m_b$ \cite{CGG}.
The aforementioned amplifications in the cut rate 
arise as corrections of the form $\Lambda/(m_b(1-x))$ and $\alpha_s 
Log^2(1-x)$.
Physically, the reasons for these enhancements are clear. The
non-perturbative corrections arise from the fact that near the
endpoint the spectrum becomes very sensitive to the Fermi
motion of the heavy quark. The logarithmic corrections are due to 
the exclusivity of the cut rate. In the limit where $x$ approaches one
there is no room for gluon radiation, thus leading to the usual
infra-red divergences of the Bloch-Nordstrom type. 

Of these two types of enhancements it is really the power corrections
that make the extraction  more difficult. The reason for this is
that the effects of Fermi motion are incalculable. In the past the
Fermi motion was modeled, leading to extractions of $V_{ub}$ for which
it was not really possible to make meaningful theoretical error
estimates. 
This is not to say that those extractions will not lead to a number
which in the end could turn out to be ``correct'', or that the old
error band is totally unreasonable. However, given that we are now entering
the age of precision CKM measurements, the theorist is 
obligated to make more systematic estimates of the  errors.
Fortunately, there is a way around having to model the Fermi motion.
The relevant point is that the effect of the Fermi motion on the
decay spectra are universal. Thus, if we can extract the necessary
information about the end point spectrum from one decay we can use
it to make a prediction for another decay spectrum.

\section{The Cut Electron Spectrum}

It can be shown from first principles that\cite{N1,BSUV}, up to corrections of order
$\Lambda/m_b$,  the
 electron energy endpoint spectrum can be written as 
\begin{equation}
\label{espect}
\frac{d\Gamma}{dE} = \int_{2E-m_b}^{\bar\Lambda}
dk_+ f(k_+) \frac{d\Gamma_p}{dE}(m_b^*),
\end{equation}
where $E$ is the charged lepton energy in semi-leptonic $B$ decay.
$m^*_b$ is the effective mass which accounts for the residual
momentum $k_+$, such that $m_b^*=m_b+k_+$   and 
the structure function, which  accounts for the Fermi motion, is
given by 
\begin{equation}
f(k_+)=\langle B(v)\mid \bar{b}_v \delta(k_+-iD_+) b_v \mid B(v)\rangle.
\end{equation}
A similar expression can be given for the radiative decay spectrum.
Thus, in principle, one could measure one end point spectrum,
deconvolve
(\ref{espect}), and then  extract $f(k_+)$ to make a prediction for the
shape of another spectrum. However, this would be a rather Herculean
task which I would wish on neither friend nor foe. Fortunately, this
analysis can be avoided by first taking the Mellin transform of the
spectrum\footnote{At tree level it can also be avoided without taking
moments \cite{N2}.}. 
The point is that 
the Mellin transform turns the convolution in (\ref{espect}) into
a product. Thus we may remove all dependence on the structure function\cite{KS}
by first taking the ratio of the moments of two spectra and then
taking the inverse Mellin transform, just as we do in relating
deep-inelastic scattering to Drell-Yan processes. This is exactly what was done
for the electron energy spectrum in ref. \cite{LLR1,LLR2}.
In this reference it was shown that following this procedure
 $|V_{ub}|^2/|V_{ts}^*V_{tb}|^2$
may be extracted from the relation \cite{LLR1,LLR2}

\begin{equation}
\label{res}
\frac{|V_{ub}|^2}{|V_{ts}^* V_{tb}|^2} =
\frac{3\,\alpha\,C_7^{(0)}(m_b)^2}{\pi}(1+H^\gamma_{mix})
 \int^1_{x_B^c}dx_B\frac{d\Gamma}{dx_B}\times\,\left\{\int^1_{x_B^c}du_B W(u_B)
   \,\frac{d\Gamma^\gamma}{du_B}\,
   \right\}^{-1} ,
\end{equation}
$H^\gamma_{mix}$ represents the corrections due to interference coming
from the operators $O_2$ and $O_8$ \cite{N2} 
\begin{equation}
\label{newfactor}
H^\gamma_{mix} = \frac{\alpha_s(m_b)}{2\pi C_7^{(0)}}\left[
C_7^{(1)} + C_2^{(0)} \Re(r_2) + 
C_8^{(0)} \left(\frac{44}9 - \frac{8\pi^2}{27}\right)\right],
\end{equation}
and $x_B^c$ is the value of the cut.
In Eq.~(\ref{newfactor}), all the Wilson coefficients,  evaluated at
$m_b$, are ``effective'' as
defined in \cite{CMM},  and 
$\Re(r_2) \approx -4.092 + 12.78(m_c/m_b - 0.29)$ \cite{GHW}.
The numerical values of the Wilson coefficients are: 
$C_2^{(0)}(m_b) \approx 1.11,\ C_7^{(0)}(m_b)\approx -0.31,\ 
C_7^{(1)}(m_b)\approx 0.48$, and $C_8^{(0)}(m_b)\approx -0.15$. 
The diagonal pieces from
$O_2$ and $O_8$ are numerically insignificant.
The function $W(u_B)$ is given by 
\begin{eqnarray}
W[u_B]=u_B^2\int^{u_B}_{x_B^c}dx_B\left(1-3(1-x_B)^2+
\frac{\alpha_s}{\pi}
(\frac{7}{2}-\frac{2 \pi^2}{9}-\frac{10}{9} Log(1-\frac{x_B}{u_B}))\right).
\end{eqnarray}

\subsection{The End Point Logs}
Over the years there has been a concern over the perturbative part of
the calculation discussed above \cite{FJMW}. This concern arises due to the
fact that as the cut approaches the end-point, the perturbative
corrections
grow. This growth is due to the  so called ``Sudakov Logs'',
$\alpha_s Log^2(1-x_c)$. These logs appear are  a consequence of the fact
that near the end-point there is limited phase space available 
for radiation, and the fact is that you can't stop radiation; you can
only hope to contain it. Indeed, it is well known that if one sums up
these double logarithms, the rate dies off as  $\Gamma \propto 
exp[-\alpha_s Log^2(1-x_c)]$. The above mentioned concern, was that
perhaps this exponentiation, and moreover, the exponentiation of
sub-leading
single logs would endanger the convergence of the perturbative
expansion. Indeed, if we assume that the logs dominate the expansion 
at each order in perturbation
theory, then we really should reorganize the calculation  in terms of
an
expansion  in the
exponent. In fact, in moment space it was shown that \cite{KS,AR} the
perturbative expansion could be written as
\begin{equation}
\frac{d\Gamma}{dN}\propto exp(Log(N) f(\alpha_s Log(N))+g(\alpha_s
Log(N))
+\alpha_s h(Log(N)) + ....).
\end{equation}
Thus the question of the 
convergence of the perturbative expansion becomes one of the nature
of the series in the exponent. Not only should the series in the
exponent converge (at least in an asymptotic sense), but it had also
better be that the first term dropped in the exponent is much less
than
one. In \cite{LLR1}, the logs were resummed including the subleading
function $g$. There it was found that, after taking the ratio of
the semi-leptonic and radiative decay rates,  
the net effect of resummation was only at $10-15\%$ level.
Part of the reason for the smallness of this effect 
is that the $f$ functions for the two
processes are identical, therefore
the leading double logs cancel in the ratio. The closed form expression
for $V_{ub}$ including the resummation is given by the same expression
as above (\ref{res}), with a new expression for the function $W$,
\begin{equation}
W[u_B]=u_B^2\int_{x_B^c}^{u_B} K\left[x_B;\frac{4}{3\pi
\beta_0}Log(1-\alpha_s \beta_0 l_{x_B/u_B})\right].
\end{equation}
\begin{eqnarray}
K(x;y) &=& 6\left\{\left[1+\frac{4\alpha_s}{3\pi}\left(1
 -  \psi^{(1)}(4+y) \right)\right]\frac1{(y+2)(y+3)}
  \right.\nonumber\\
&&\phantom{6\{}- \frac{\alpha_s}{3\pi}
  \left[\frac{1}{(y+2)^2}-\frac{7}{(y+3)^2}\right]
 -\left.\frac{4\alpha_s}{3\pi}\left[ \frac1{(y+2)^3}-\frac1{(y+3)^3}\right]
  \right\} \nonumber\\
&& - 3(1-x)^2.
\end{eqnarray}
The function $K$ has a non-integrable singularity
 in it due the fact that the sum in the
exponent should be considered in terms of  a prescription for a non-Borel
summable
series. Due to its asymptotic nature it can be well
approximated by expanding the argument of the Log in a series and
keeping as many terms as one wishes until the series starts to
diverge.
An excellent
approximation for $W$ is given by expanding the second argument of 
$K$\cite{LLR4},
\begin{equation} \frac{4}{3 \pi \beta_0}Log(1-\alpha_s \beta_0 l_{x_B/u_B})\approx
\alpha_s \frac{4}{3 \pi}Log(1-x_B/u_B)-\alpha_s^2\frac{25}{18 \pi^2}
Log^2(1-x_B/u_B). 
\end{equation}
While the use of this improved prediction may shift the central value
slightly, it will not change the error bars.

\section{The Hadronic Mass Cut}
It is also possible to remove the background from 
charmed transitions by cutting on the hadronic invariant mass.
While this choice presents a greater experimental challenge, it 
benefits from the fact that, unlike the electron spectrum,
most of the $B\rightarrow X_u e \nu$ decays are expected to lie within
the region $s_H<M_D^2$. 
Furthermore, it is believed that 
even though both the invariant mass region  $s_H<M_D^2$
and electron energy regions $M_B/2>E_e>(M_B^2-M_D^2)/(2 M_B)$ receive
contributions from hadronic final states with invariant mass up 
to $M_D$, the cut mass spectrum will be less sensitive
to local duality violations.  This belief rests on the fact that  
 the contribution of
large mass states is kinematically suppressed 
for the  electron energy spectrum in the region of interest.
The expression for $V_{ub}$ for the case of the hadronic mass cut
is given by \cite{LLR3}
\begin{equation}
\frac{|V_{ub}|^2}{|V_{ts}|^2}=
\frac{6\,\alpha\,C_7(m_b)^2(1+H_{mix}^\gamma)\,\delta\Gamma(c)}
{\pi\,[I_0(c)+I_+(c)]}.
\end{equation} 
The expressions for $\Gamma(c),~I_0(c)$ and $I_+(c)$ can be found in
\cite{LLR3}. The leading errors in this expression are again of order
$~\Lambda/m_b$.
The effect of resummation of the end-point logs in this case was shown
to be negligible \cite{LLR2}.

\section{Some Caveats and Concluding Remarks}
We have emphasized the fact that we are now capable of extracting
$\mid V_{ub} \mid$ in a systematic fashion. Which is to say that we have our
errors under control. The leading errors are of order $\Lambda/m_b$
and $\alpha_s (1-x)$. However, strictly speaking these are really
the only errors we know how to quantify. As I emphasized the
calculation
is based upon certain assumptions about local duality. We don't know
how to quantify these errors, all we know is when we do and do not
expect local duality to be a valid assumption. Basically, the point is that
we expect hadronic observables to well approximated in terms of
partonic calculations when we are able to smear over resonances 
in a ``sufficient manner'' \footnote{Mathematically this assumption
boils down to the fact that we are doing an operator product expansion
in a kinematic region where its not really justified.}.
If the final state phase space is so restrictive that we only
average over one or two resonances then, we begin to worry that
our assumption is not well founded. 
It is interesting to note that in the charmed decay, the inclusive
rate is saturated up to corrections of order $\Lambda^2/m^2$
by including only the $D$ and $D^*$ in the final states\cite{BGM} in
the small velocity limit \cite{SV2}.
Naively, one would think that pion emission would only be suppressed
by $p_\pi/f_\pi$, but in the zero recoil limit the currents become
generators of the effective theory and can only produce linear
combinations
of the $D$ and $D^*$. While this is tantalizing, and gives one hope
that local duality will work well even with small numbers of
resonances, this case  may be considered special.

None the less, the best  choice of cuts, as far
as local duality is concerned, is the hadronic mass, as it is
believed to contain approximately $80\%$ of the total rate, and if
duality is going to work somewhere this is the place. Indeed, given
that this is a relatively large percentage of the rate, compared to
those expected in the electron energy cut $\simeq 20\%$ or leptonic
mass cut $\simeq 20\%$ (which does need the radiative decay data
\cite{BLL1,BLL2,B}), we may hope to actually test the duality 
assumption. This may be done by simply varying the cut and seeing
if the extracted value of $V_{ub}$ remains fixed. 

Let me pause at this point to make a few comments regarding some
issues which were raised during the conference. First of all, it
is often asked, ``if we know that the hadronic mass cut really
does include $80\%$ of the rate, can't we get a relatively good 
 extraction of
$V_{ub}$ without care about the radiative decays?''. The answer to
this is that we really don't know that the hadronic mass cut captures
$80\%$ of the rate. That number is a rough estimate which is made using
an inspired guess for the wave function. The purpose of the number is
simply to get a feeling for relative merits of the various cuts, but
since the whole point of the exercise is to eliminate model
dependence, these percentages should not be taken too seriously.
Secondly I would like to make a few comments about the recent CLEO 
extraction, as discussed in Roy Brieres' talk at this conference
\cite{Roy}. As I understand it, 
the extraction was performed by using a guess for the
structure function, and varying the parameters of the model until a
fit to the radiative decay spectrum was found. This fit was then used to make
a prediction for the cut rate in the semi-leptonic decay.
The robustness of this method was then ``tested'' by using two
different parameterizations for the structure function. 
Now, I believe this is not the best way to do things. There are, no
doubt, a large number of parameterizations for the structure function 
which could fit the
radiative decay data, and given that the function with which this
parameterization
is convoluted differs in the two different decay modes, there is no
reason that the results should necessarily be identical in 
any parameterization. Of course, even in the method I described in
this talk one still needs to fit the radiative decay data, and this
may indeed be well fit by using a model. The
point is that in the method described above, all that is need is the
{\it physical spectrum}, it doesn't matter what model you use to fit the
data. In the end it may well indeed be that both methods give the same
answer, but the present CLEO extraction as it stands, is not justified
from first principles.

\begin{theacknowledgments}
I would like that thank my collaborators on this subject, Adam
Leibovich
and Ian Low. Thanks also go the organizers of this conference who
were able to pull off a constructive and pleasant conference during
difficult times. This work was supported in part by the Department of Energy 
under grant number DOE-ER-40682-143 and DE-AC02-76CH03000.
\end{theacknowledgments}


\doingARLO[\bibliographystyle{aipproc}]
          {\ifthenelse{\equal{\AIPcitestyleselect}{num}}
             {\bibliographystyle{arlonum}}
             {\bibliographystyle{arlobib}}
          }
\bibliography{writeup}

\hyphenation{Post-Script Sprin-ger}
\begin{thebibliography}{20}
\expandafter\ifx\csname natexlab\endcsname\relax\def\natexlab#1{#1}\fi
\providecommand{\enquote}[1]{``#1''}
\expandafter\ifx\csname url\endcsname\relax
  \def\url#1{\texttt{#1}}\fi
\expandafter\ifx\csname urlprefix\endcsname\relax\def\urlprefix{URL }\fi

\bibitem[Chay et~al.(1990)]{CGG}
Chay, J., Georgi, H., and Grinstein, B., \emph{Phys. Lett.}, \textbf{B247},
  399--405 (1990).

\bibitem[Shifman and Voloshin(1985)]{SV1}
Shifman, M.~A., and Voloshin, M.~B., \emph{Sov. J. Nucl. Phys.}, \textbf{41},
  120 (1985).

\bibitem[Neubert(1994{\natexlab{a}})]{N1}
Neubert, M., \emph{Phys. Rev.}, \textbf{D49}, 3392--3398 (1994{\natexlab{a}}).

\bibitem[Bigi et~al.(1994)]{BSUV}
Bigi, I. I.~Y., Shifman, M.~A., Uraltsev, N.~G., and Vainshtein, A.~I.,
  \emph{Int. J. Mod. Phys.}, \textbf{A9}, 2467--2504 (1994).

\bibitem[Neubert(1994{\natexlab{b}})]{N2}
Neubert, M., \emph{Phys. Rev.}, \textbf{D49}, 4623--4633 (1994{\natexlab{b}}).

\bibitem[Korchemsky and Sterman(1994)]{KS}
Korchemsky, G.~P., and Sterman, G., \emph{Phys. Lett.}, \textbf{B340}, 96--108
  (1994).

\bibitem[Leibovich et~al.(2000{\natexlab{a}})]{LLR1}
Leibovich, A.~K., Low, I., and Rothstein, I.~Z., \emph{Phys. Rev.},
  \textbf{D61}, 053006 (2000{\natexlab{a}}).

\bibitem[Leibovich et~al.(2000{\natexlab{b}})]{LLR2}
Leibovich, A.~K., Low, I., and Rothstein, I.~Z., \emph{Phys. Rev.},
  \textbf{D62}, 014010 (2000{\natexlab{b}}).

\bibitem[Chetyrkin et~al.(1997)]{CMM}
Chetyrkin, K., Misiak, M., and Munz, M., \emph{Phys. Lett.}, \textbf{B400},
  206--219 (1997).

\bibitem[Greub et~al.(1996)]{GHW}
Greub, C., Hurth, T., and Wyler, D., \emph{Phys. Lett.}, \textbf{B380},
  385--392 (1996).

\bibitem[Falk et~al.(1994)]{FJMW}
Falk, A.~F., Jenkins, E., Manohar, A.~V., and Wise, M.~B., \emph{Phys. Rev.},
  \textbf{D49}, 4553--4559 (1994).

\bibitem[Akhoury and Rothstein(1996)]{AR}
Akhoury, R., and Rothstein, I.~Z., \emph{Phys. Rev.}, \textbf{D54}, 2349--2362
  (1996).

\bibitem[Leibovich et~al.(2001)]{LLR4}
Leibovich, A.~K., Low, I., and Rothstein, I.~Z., \emph{Phys. Lett.},
  \textbf{B513}, 83--87 (2001).

\bibitem[Leibovich et~al.(2000{\natexlab{c}})]{LLR3}
Leibovich, A.~K., Low, I., and Rothstein, I.~Z., \emph{Phys. Lett.},
  \textbf{B486}, 86--91 (2000{\natexlab{c}}).

\bibitem[Boyd et~al.(1996)]{BGM}
Boyd, C.~G., Grinstein, B., and Manohar, A.~V., \emph{Phys. Rev.},
  \textbf{D54}, 2081--2096 (1996).

\bibitem[Shifman and Voloshin(1988)]{SV2}
Shifman, M.~A., and Voloshin, M.~B., \emph{Sov. J. Nucl. Phys.}, \textbf{47},
  511 (1988).

\bibitem[Bauer et~al.(2000)]{BLL1}
Bauer, C.~W., Ligeti, Z., and Luke, M.~E., \emph{Phys. Lett.}, \textbf{B479},
  395--401 (2000).

\bibitem[Bauer et~al.(2001)]{BLL2}
Bauer, C.~W., Ligeti, Z., and Luke, M.~E., \emph{Phys. Rev.}, \textbf{D64},
  113004 (2001).

\bibitem[Bauer(2001)]{B}
Bauer, C.~W., \emph{These proceedings} (2001).

\bibitem[Briere(2001)]{Roy}
Briere, R., \emph{These proceedings} (2001).

\end{thebibliography}

\end{document}